\documentstyle[aps,preprint,epsfig]{revtex}

\begin{document}

\title{Exact solutions of Einstein equations }

\author{A.C.V.V. de Siqueira\footnote{acvvsiq@hotlink.com.br}}
\address{Departamento de F\'{i}sica e Matem\'{a}tica - Universidade Federal Rural de
Pernambuco, 52171-900, Recife, PE, Brazil}

\date{\today }
\maketitle

\begin{abstract}
We use a metric of the type Friedmann-Robertson-Walker to
obtain new exact solutions of Einstein equations for a scalar and massive
field.The solutions have a permanent or transitory inflationary behavior.
\end{abstract}


\newpage
\section{Introtuction}
There is a great interest in the solutions of Einstein equations due to
their sucess in many of the areas of the physics. Exact solutions make
possible a better knowledge of the many problems in several fields,
like, for example, Kerr's solution in Astrophysics or Robertson-Walker in
Cosmology \cite{Hawking}. The new solutions presented in this paper have an
initial or permanent behavior of inflationary nature. Inflation is a
very necessary scenario for a primordial universe \cite{Guth}. This paper
is organized as follows. In section II we introduce the equations of
Einstein-Klein-Gordon and present their solutions. In section III we
present our final considerations.

\section{ The Einstein's Field Equations and their solutions}
We use a local basis, with the components of Riemann tensor given by,

\begin{equation}
R^{\alpha}{}_{\mu \sigma \nu }=\partial_{\nu} \Gamma_{\mu \sigma
}^{\alpha}-\partial_{\sigma}\Gamma_{\mu \nu}^{\alpha}+\Gamma_{\mu
\sigma}^{\eta}\Gamma_{n \nu}^{\alpha}-\Gamma_{\mu \nu}^{\eta}\Gamma
_{\sigma \eta}^{\alpha} \text{,} \label{b1}
\end{equation}

and components of Ricci tensor

\begin{equation}
R_{\mu \nu}=R^{\alpha}_{\mu \alpha \nu } \text{.} \label{b2}
\end{equation}

The Einstein`s equations, with cosmological constant $\Lambda $, are

\begin{equation}
R_{\mu \nu }-\frac{1}{2}g_{\mu \nu }R+\Lambda g_{\mu \nu}=-\frac{8\pi G}{
c^{2}}T_{\mu \nu} \text{,} \label{b3}
\end{equation}

where $T_{\mu \nu}$ is the tensor of momentum-energy for a massive
and scalar field \cite{Lopes},

\begin{equation}
T_{\mu \nu }=2 \nabla_{\mu} \phi \nabla_{\nu} \phi
-g_{\mu \nu}\nabla^{\alpha } \phi \nabla_{\alpha}\phi +m^{2}
g_{\mu \nu }\phi^{2} \text{.} \label{b4}
\end{equation}

The massive scalar field obeys to the following motion equation

\begin{equation}
\partial _{\mu \text{ }}\{\sqrt{-g}g^{\mu \nu }\partial_{\nu }\phi
\}+m^{2}\sqrt{-g}\phi =0 \text{.} \label{b5}
\end{equation}

We use (+, -, -, -) signature convention and one line element of the type
Friedmann-Robertson-Walker, given by

\begin{equation}
ds^{2}=dt^{2}-\frac{d\sigma^{2}e^{g}}{\left( 1+Br^{2}\right)^{2}}
\label{b6}
\end{equation}

where $d \sigma^{2}$ is the three-dimensional Euclidian line element.
In this paper we use $A=8\pi G/c^{2}$, $c=1$, $B=k/4a^{2}$ and with $ k=0$ ,
$k=1$ and $k=-1$. We also have $a^{2}$ a constant.

Due to isotropy and homogeneity of (\ref{b6}), the equations (\ref{b3}) and (\ref
{b5}) assume the following forms

\begin{equation}
\left( \frac{d\phi }{dt}\right) ^{2}+m^{2}\phi ^{2}+\frac{\Lambda
}{A}-\frac{3}{4A}\left( \frac{dg}{dt}\right) ^{2}-\frac{12B}{A}e^{-g}=0
\text{,} \label{b7}
\end{equation}

\begin{equation}
\left(\frac{d\phi }{dt}\right)^{2}-m^{2}\phi^{2}-\frac{\Lambda}{A}+\frac{3}
{4A}\left(\frac{dg}{dt}\right)^{2}+\frac{1}{A}\frac{d^{2}g}{dt^{2}}+
\frac{4B}{A}e^{-g}=0  \label{b8}
\end{equation}

and

\begin{equation}
\frac{d^{2}\phi }{dt^{2}}+\frac{3}{2}\frac{dg}{dt}\frac{d\phi }{dt}
+m^{2}\phi=0  \label{b9}
\end{equation}

where $\phi =\phi \left( t\right) $ and (\ref{b9}) is the motion equation of
the field. We will present now the three exact solutions of the system formed
by the equations (\ref{b7}), (\ref{b8}) and (\ref{b9}). We consider, initially, the
case $B=0$, followed by $B>0$ and later by $B<0$.\\
 For $B=0$ the field is given by

\begin{equation}
\phi =\frac{\in mt}{\sqrt{3A}}+b  \label{b10}
\end{equation}

with $\in =\pm 1$ and $b$ an arbtrary constant.

In this case the cosmological constant obeys the condition

\begin{equation}
\Lambda =-\frac{m^{2}}{3},  \label{b11}
\end{equation}

a negative value, associated with the mass of the scalar field.
The corresponding line element will be

\begin{equation}
ds^{2}=dt^{2}-d\sigma ^{2}e^{[-2\in mb(\sqrt{\frac{A}{3}})t-\frac{m^{2}}{3}
t^{2}]}, \label{b12}
\end{equation}

that, for special conditions has a transitory inflationary nature followed
by a contration as we will see in section III$ $.

 Let us consider the condition $B>0$. In this case the cosmological
constant $\Lambda $ is proportional to the square of the mass, but
is a positive constant, given by,

\begin{equation}
\Lambda =\frac{3m^{2}}{2} \text{.} \label{b13}
\end{equation}

For the field, we obtain the following result,

\begin{equation}
\phi =\frac{\in }{m}\sqrt{\frac{8B}{A}}e^{(-\frac{\in mt}{\sqrt{2}})}
\label{b14}
\end{equation}

The line element of the metric will be,

\begin{equation}
ds^{2}=dt^{2}-d\sigma ^{2}\frac{e^{(\in \sqrt{2}mt)}}{(1+Br^{2})^{2}}
\text{.} \label{b15}
\end{equation}

Let us see now $B<0$.

In this case we have to be more careful in the analyse of the results
having it in mind, we obtain, for the the field, the expression

\begin{equation}
\bigskip \phi =\in \sqrt{\frac{12B}{\Lambda A}}e^{(-\frac{\in mt}{\sqrt{2}})}
\label{b16}
\end{equation}

The cosmological constant $\Lambda $, for the above conditions, will also be
given by (\ref{b13}). Substituting (\ref{b13}) in (\ref{b16}) the field assume the
following form,

\begin{equation}
\phi =\in \sqrt{\frac{8B}{Am^{2}}}e^{(-\frac{\in mt}{\sqrt{2}})}, \label{b17}
\end{equation}

and the correspondent line element will be,

\begin{equation}
ds^{2}=dt^{2}-d\sigma^{2}\frac{e^{(\in \sqrt{2}mt)}}{(1+Br^{2})^{2}}.
\label{b18}
\end{equation}

\section{Conclusions}

We will now make some considerations about the results of the
previous section. In the case $B=0$ will specialize the constants $\in $ and $b$ so
that we have an initial inflationary phase. Let us take $\in =-1$ and $b>0$.
The line element (\ref{b12}) will now have a dominant inflationary phase to
$t<\left( 6b/m\right) \sqrt{A/3}$, followed by a contraction for $t>\left(
6b/m\right) \sqrt{A/3}$. The field, given by (\ref{b10}), will be positive for
$t<\left( b/m\right) \sqrt{3A}$ and negative for $t>\left( b/m\right) \sqrt{
3A}$. If we assume that $\in =+1$ and $b<0$, the line element will not
change and will have an initial dominant inflationary phase to $t<-\left(
6b/m\right) \sqrt{A/3}$ and one contracted phase
to $t>-\left( 6b/m\right) \sqrt{A/3}$. In this two situations, as it can be
seen by motion equation (\ref{b9}), the field will not have
a dynamic character, but associated with the cosmological constant $\Lambda
$, both will have a decisive dynamic character and they suggest that their
interactions with the other fields of the nature can be thought as fields in
a background field. As a consequence, the space-time of the metric (\ref{b12})
can be thought as a reservoir of energy, and as an infinite arena where the other
fields of nature interact \cite{Zel'dovich}. For $B>0$, we will have a finite
space-time, with an inflationary expansion that coud be weakened by a
processes of mass generation of the gauge fields, and greatly slowed down
by the gravitational interaction with the cosmic fluid. It is interesting
to observe that in the present days the value of $\Lambda $ is estimated
in $10^{-54}$ cm$^{-2}$. This means that the mass of the field is very small
 \cite{See}. The expansion would be very slow and we could think about particles
confinated in a background field with topology R$\times $S$^{3}$.
Let us now consider the case $B<0$. With this condition we will have an
imaginary field, given by (\ref{b17}) in an infinite universe (\ref{b18}), since
the mass of the field was real and positive. If the mass is considered as an
imaginary constant the field and the metric (\ref{b18}) will are complex.
In the two situations we will have problems with the momentum-energy tensor.
We conclude this paper admiting that part of this section is speculative
because we have not made applications of our results, but in some
cases, it will be possible to accomphish estimates and confront them with
the observational data, and conclud if the new solutions are or not physically
useful.


\begin{thebibliography}{99}

\bibitem{Hawking}  S. W. Hawking and G. F. R. Ellis, \emph{The Large Scale Structure of
Space-Time} (Cambridge University Press, 1973); V. P. Frolov and I. D. Novikov,
\emph{Black Hole Physics-Basic Concepts and New
Developments} (Kluwer Academic Publishers, 1998); S. Weinberg,
\emph{Gravitations and Cosmology: Principles and
Applications of the General Theory of Relativity} (John Wiley Sons,1972).

\bibitem{Guth}  A. H. Guth, Phys. Rev. D \textbf{23}, 347 (1981).

\bibitem{Lopes}  J. L. Lopes, \emph{Gauge Field Theories, An Introduction} (Pergamon
Press, 1981); A. A. Grib, S. G. Mamayev and V. M. Mostepanenko, \emph{Vacuum Quantum Effects in
Strong Fields} (Friedmann Laboratory Publishing, St. Petersburg, 1994).

\bibitem{Zel'dovich}  Ya. B. Zel'dovich, Sov. Phys. \textbf{USPEKHI, vol. 11}, 1968.

\bibitem{See}  See ref. [4], p. 382.

\end{thebibliography}
\end{document}